# Tachyons, Quanta and Chaos

By

Mark Davidson

February 15, 2001

Spectel Research Corporation
807 Rorke Way
Palo Alto, CA  94303
email: mdavid@spectel.com

## Abstract

It is shown using numerical simulation that classical charged tachyons have several features normally thought to be unique to quantum mechanics.  Spin-like self-orbiting helical motions are shown to exist at discrete values for the velocity of the tachyon in Feynman-Wheeler electrodynamics and in normal causal electrodynamics more complex closed orbits also appear to exist.   Tunneling behavior of the classical tachyon is observed at classical turning points depending on the angle of incidence.  The equations of motion appear to be chaotic and effectively indeterministic when the tachyon crosses its own past light cone.  It is argued that self-interacting tachyons moving in a tight helix would behave causally, and that they could be a basis for a hidden variable description of quantum mechanics.    A procedure is proposed which could determine the fine structure constant.

## 1.  Introduction

Arnold Sommerfeld first published the possibility of tachyons in 1904 [1], at the same time as the young Albert Einstein was neglecting his chores in the Swiss patent office in order to revolutionize the world of physics.  Tachyons captured some interest in the physics community in the 1960s and 70s [1-7], but they have since fallen somewhat from fashion because direct experimental evidence has not been found to support their existence, and also because of concerns about causality [8].   Arguments have been made to counter the causality objections [9], and the issue remains in dispute.  There are several reasons why tachyons are still of interest today, and in fact interest may be increasing.  First, many string theories have tachyons occurring as some of the particles in the theory [10], although they are generally regarded as unphysical in those theories.   There are also several recent papers that assert experimental evidence that some neutrinos are tachyons [11,12].  There is a new and extensive re-analysis of tachyon dynamics [13].  There is much discussion in the physics literature in recent years of superluminal connections implied by quantum mechanics and by the evanescent wave phenomenon of light optics as well as quantum tunneling, all indirect evidence of nonlocality in nature. These and other recent developments show that tachyons are still a timely subject for investigation.

Tachyon trajectories can conceivably intersect their own past and future light cones [4], and if they were charged they would experience an electromagnetic self-force that has no analog for particles moving slower than light. It is the exploration of this self-interaction that is the subject of this paper. If a tachyon were moving very fast, then it could intersect with its own light cone at a large number of points in its own past or future. The resulting theory looks to be extremely complex and fertile for investigation, having more in common with a many body problem than with a classical single particle trajectory problem. According the analysis of Ey and Hurst [5], charged tachyons will not experience a local self force (the type that occurs in the Lorentz-Dirac Equation). Their arguments rely on a regularization procedure that seems quite natural and reasonable. The tachyons will experience a self force when they cross their own past light cone as we show here, and in this case they can radiate at least in the usual causal electrodynamics with retarded potentials. In the Feynman-Wheeler formulation of electrodynamics they will not radiate.

Recami [6] has presented a thorough analysis of the various possibilities for classical equations of motion for tachyons. It is shown here that in Feynman-Wheeler action at a distance electrodynamics these equations have many closed circular or helical orbits, depending on the speed of the tachyon in the circular orbit. Some speeds give bound orbits while others do not. The self-orbiting tachyon thus appears to be moving slower than the speed of light on the average, and to have an intrinsic angular momentum. One might intuitively expect that a tachyon that is moving in a circle so fast as to see its own image – ie. to intersect its own light cone – would be repelled from its own image since like charges repel. Although this is usually true, the force can sometimes become attractive because magnetic and relativistic effects must be included which lead to net attraction in some cases, depending on the speed of the orbit as numerical simulations show.

First the Feynman-Wheeler form for Electrodynamics is considered and it is shown that helical solutions exist. Then causal electrodynamics is considered and the azimuthal force that causes the motion to be non-helical is calculated. The radial force for helical motion in this case is the same as for Feynman-Wheeler interaction. The calculations are too complicated to perform analytically, and so numerical computations are presented.

## 2. Tachyon Equations of Motion

The equations of motion for classical tachyons have been developed by many authors. The most commonly accepted form for the Tachyon equations of motion are (Recami[6], in units where c = 1):

$$\mathbf{F} = \frac{d}{dt}(\frac{m_0 \mathbf{V}}{\sqrt{\mathbf{V}^2 - 1}}) \tag{1}$$

Recami argues that the force equation could with equal plausibility be chosen to be (see section 6'14 in Reference 6)

$$\mathbf{F} = -\frac{d}{dt}(\frac{m_0 \mathbf{V}}{\sqrt{\mathbf{V}^2 - 1}}) \tag{2}$$

where in both of these equations the mass $m_0$ is positive and real. The two equations differ by an overall sign. These equations are completely different and have very different solutions. For the remainder of this paper we shall use only the form in equation 1.

We shall first illustrate the solution for the time symmetric action at a distance electrodynamics of Fokker, Feynman, and Wheeler [14-15]. The slower than light problem of a relativistic two-body bound state has been solved for this type of interaction [16] which lends encouragment toward looking for a solution in this context in the present case.

## 3. Action-at-a-Distance Electrodynamics

When the force is due solely to electromagnetic effects, one may write for equation 1 of motion of the electrically charged tachyon:

$$\frac{d}{dt}(\frac{m_0}{\sqrt{\mathbf{v}^2/c^2 - 1}}\frac{d\mathbf{x}}{dt}) = q[\mathbf{E} + \frac{\mathbf{v}}{c} \times \mathbf{B}] \tag{3}$$

where $q$ is the charge of the tachyon and

$$\mathbf{E} = -\nabla\Phi - \frac{1}{c}\frac{d\mathbf{A}}{dt}; \quad A^\mu = (\mathbf{A}, i\Phi) \tag{4}$$

and where

$$A^\mu = \frac{1}{c}\iint\frac{J^\mu(\mathbf{x}',t')}{R}[\frac{1}{2}\delta(t'+\frac{R}{c}-t) + \frac{1}{2}\delta(t'-\frac{R}{c}-t)]d^3x'dt', R = |\mathbf{x} - \mathbf{x}'| \tag{5}$$

where the currents $\boldsymbol{J}$ are due to all the relevant charges. The expression (5) is symmetrical between the future and the past since the advanced and retarded conditions enter with equal weights in the Feynman-Wheeler approach..

In (5) must be included the self interaction possibilities since the tachyon moves faster than light and it can therefore interact through the retarded and advanced potential with its own trajectory. Thus in (5), *J* includes not only the current density produced by all the other relevant particles, but also that of the particular particle whose motion the equation describes. Excluded from *J*, however, is the singular contribution arising from the particle's present location, just as this is also excluded in the slower than light case.

Next we consider an isolated electrically charged tachyon. The only contributions to *J* in equation 5 is then from the particle's own trajectory intersecting its own light cone.

For any particle, tachyon or tardyon, which moves strictly forward in time we may write the contribution to *J* as

$$J^{\mu}(\mathbf{x},t) = qc\beta^{\mu}\delta^3(\mathbf{x}-\mathbf{r}(t)), \quad \beta^{\mu} = (\frac{1}{c}\frac{d\mathbf{x}}{dt},i) = (\boldsymbol{\beta},i) \tag{6}$$

The advanced and retarded potentials can be written

$$A^{\mu}_{\substack{Ret \\ Adv}}(\mathbf{x},t) = q\int \frac{\beta^{\mu}(t')}{|\mathbf{x}-\mathbf{x}(t')|}\delta(t'\pm\frac{|\mathbf{x}-\mathbf{x}(t')|}{c}-t)dt' \tag{7}$$

Let us use the notation

$$\mathbf{R}(\mathbf{x},t') = \mathbf{x}-\mathbf{x}(t'); \quad R(\mathbf{x},t') = |\mathbf{R}(\mathbf{x},t')| \tag{8}$$

So that **R** is the vector pointing from the source point on the particle trajectory to a test point **x**. Then defining

$$K_{\substack{Ret \\ Adv}}(\mathbf{x},t') = \frac{d}{dt'}(t'\pm\frac{R(\mathbf{x},t')}{c}) = 1\mp\hat{\mathbf{n}}\cdot\boldsymbol{\beta} \quad \text{where } \hat{\mathbf{n}} = \frac{\mathbf{R}(x,t')}{R(x,t')} \tag{9}$$

we can write

$$\Phi^{\substack{Ret \\ Adv}} = e\left[\frac{1}{KR}\right]_{\substack{ret \\ adv}} \quad ; \quad \mathbf{A}^{\substack{Ret \\ Adv}} = e\left[\frac{\boldsymbol{\beta}}{KR}\right]_{\substack{ret \\ adv}} \tag{10}$$

where β, K and R are evaluated at the retarded or advanced time as determined by the case. The gradient operator acting on any of these fields can be expressed in terms of the R derivative

$$\nabla \Rightarrow (\nabla R)\frac{\partial}{\partial R} = \hat{\mathbf{n}}\frac{\partial}{\partial R} \tag{11}$$

Then we find the following expressions for electric and magnetic fields analogous to the slower than light case:

$$\mathbf{E}_{adv}^{ret} = \left[ q\frac{(\hat{\mathbf{n}} \mp \boldsymbol{\beta})}{K^3 R^2}\left(1 - \beta^2\right) + \frac{q}{cK^3 R}\hat{\mathbf{n}} \times \left((\hat{\mathbf{n}} \mp \boldsymbol{\beta}) \times \dot{\boldsymbol{\beta}}\right) \right]_{\substack{ret \\ adv}} \tag{12}$$

and

$$\mathbf{B}_{adv}^{ret} = \pm \hat{\mathbf{n}}_{\substack{ret \\ adv}} \times \mathbf{E}_{adv}^{ret} \tag{13}$$

Notice an interesting feature of equation 12. K as defined in equation 9 can be zero. Whenever K vanishes, the electric field in (12) can be singular even if R is not zero. This singular behavior has no analogue in slower than light electrodynamics. The vanishing of K defines the Cerenkov cone, and singularities on this cone were also noted in [5].

## 4. Circular Motion Solutions

We now look for solutions to the equations of motion (1) in the form of circular motion:

$$x(t) = r\cos(\omega t); \quad y(t) = r\sin(\omega t); \quad z = 0 \tag{14}$$

and where the speed is superluminal

$$v = r\omega > c \tag{15}$$

The helical trajectory may instersect the lightcone of a testpoint on the trajectory at a number points as is illustrated in Figure 1. The number of intersections is a function only of the particle's speed. We must sum the force contribution from all of these intersection points. The points come in symmetrical pairs, each retarded point being the reflection of the correspondng advanced point.

Depending on how fast the tachyon is moving, the trajectory will intersect the future and past light cones symmetrically at least at one point. Consider Figure 2 as we calculate the fields at an arbitrary test point along the trajectory which are due to a pair of light cone intersections. The total field at the test point is given by:

$$\mathbf{E} = \frac{1}{2}\left(\mathbf{E}_{ret} + \mathbf{E}_{adv}\right), \quad \mathbf{B} = \frac{1}{2}\left(\mathbf{B}_{ret} + \mathbf{B}_{adv}\right) \tag{16}$$

For circular motion $\dot{\boldsymbol{\beta}}$ points in the negative radial direction in Figure 2. Notice from figure 2 and equation 9 that for this case of reflection pairs of intersecting points that

$$K_{ret} = K_{adv} \equiv K \tag{17}$$

Neither the retarded nor the advanced force is in the radial direction. However, it is straightforward to show that when one adds the contribution of a pair of advanced and retarded symmetrical intersection points, that the force acting on the test point due to this contribution becomes a radial force. Schild [16] noted this same fact for slower than light circular orbiting particles.

We have solved the force problem numerically, and we find the following result for the radial force acting on the test point:

$$F_r = -\frac{q^2}{r^2} Z(\beta) \tag{18}$$

where Z is a dimensionless function depending only on $\beta$ and where positive values of Z correspond to attraction to the center of the helix. Notice in (12) that since the acceleration varies as $1/r$ the electric field will have an overall $1/r^2$ dependence leading to (18). A plot of Z is presented in Figure 3 up to $\beta = 20$. Whenever Z is positive a helical motion solution exists to (1). When Z is negative, a helical solution exists for equation (2).

The equations of motion become simply:

$$\frac{m_0}{\sqrt{\mathbf{v}^2/c^2 - 1}} \frac{\mathbf{v}^2}{r} = \frac{q^2}{r^2} Z(\beta) \Rightarrow r = \frac{\sqrt{\beta^2 - 1}}{m_0 c^2 \beta^2} q^2 Z(\beta) \tag{19}$$

Notice that for certain values of $\beta$ that Z becomes discontinuous in Figure 3. These discontinuities occur when the Cerenkov cone from the source point intersects the test point. The next section explains this behavior.

## 5. An Explanation Of Why Certain Discrete Velocities Lead To An Attractive Force

The numerical simulations show singular discontinuous complex behavior in the radial and azimuthal forces when the velocity has certain discrete values. The occurrence of these singularities can be understood by the following simple geometrical analysis. Consider the retarded fields produced by the circular motion of (14). The retardation condition which determines the times t' (sourcepoints) on the trajectory that can intersect with the position of the particle at time t (testpoint) is

$$|x(t) - x(t')| = c(t - t'), \quad \text{where } t > t' \tag{20}$$

Owing to the circular form of the motion, this condition may be written as

$$r^2 \left| e^{i\omega(t-t')} - 1 \right|^2 = c^2(t - t')^2, \ t > t' \tag{21}$$

further defining
$$\beta = \omega r/c \quad \text{and} \quad \tau = c(t-t')/r \tag{22}$$

the null condition becomes

$$2 - 2\cos(\beta\tau) = \tau^2, \ \tau > 0 \tag{23}$$

The larger that $\beta$ is the more solutions there are to this equation. Defining:

$$f(\tau, \beta) = 2 - 2\cos(\beta\tau) - \tau^2, \quad \tau > 0 \tag{24}$$

Then the null points are determined by the equation

$$f(\tau, \beta) = 0, \ \tau > 0 \tag{25}$$

Noticing that the following properties are true:

$$f(0, \beta) = 0 \tag{26}$$

$$f(\tau, \beta) \approx \tau^2(\beta^2 - 1), \text{ for small } \tau \tag{27}$$

$$f(\propto) < 0 \tag{28}$$

It follows from these results that for $\beta > 1$ there is at least one solution for $\tau$ which satisfies the null condition. Now define:

$$N(\beta) = \text{The number of solutions to } f = 0 \text{ for positive } \tau \tag{29}$$

This function starts off at 1 and increases discontinuously by jumps of two at every point β at which $f$ has a solution to the following two simultaneous equations:

$$f(\tau,\beta) = 0; \quad \frac{\partial f(\tau,\beta)}{\partial \tau} = 0 \tag{30}$$

These are equivalent to:

$$\varphi = \beta\tau \tag{31}$$

$$2 - 2\cos(\varphi) - \varphi\sin(\varphi) = 0 \tag{32}$$

$$\beta = \sqrt{\varphi/\sin(\varphi)} \tag{33}$$

Equation 32 is a single transcendental which can be solved easily by numerical means. Table 1 shows the first 15 values of β:

| | | |
|---|---|---|
| 4.603338848751701e+000 | 2.039583252184294e+001 | 3.611446976533017e+001 |
| 7.789705767492714e+000 | 2.354070189773618e+001 | 3.925717095448966e+001 |
| 1.094987986982622e+001 | 2.668479810180271e+001 | 4.239970774262564e+001 |
| 1.410169533046915e+001 | 2.982836607105987e+001 | 4.554211418676631e+001 |
| 1.724976556755881e+001 | 3.297155711433862e+001 | 4.868441554248154e+001 |

Table 1  Singular β values

Next it will be proven that at $K_{\text{ret}}$ vanishes at these singular values where the number of roots to the null equation changes discontinuously by increments of 2.   This also causes the Force to be singular in a neighborhood of these points.  The velocity of the source-point is given by:

$$\beta_x(t') = -r\omega\sin(\omega t'); \quad \beta_y = r\omega\cos(\omega t') \tag{34}$$

and  $K_{\text{ret}}$ from (9) is given by

$$K_{ret}(t,t') = 1 - \boldsymbol{\beta}(t') \cdot (\mathbf{x}(t) - \mathbf{x}(t'))/|\mathbf{x}(t) - \mathbf{x}(t')| \tag{35}$$

which can be rewritten as

$$K_{ret}(t,t') = 1 - \frac{r\omega}{c}\text{Im}(e^{i\omega(t-t')} - 1)/\left|e^{i\omega t} - e^{i\omega t'}\right| \tag{36}$$

or as

$$K_{ret}(t,t') = 1 - \beta \sin(\varphi) / \sqrt{2 - 2\cos(\varphi)}$$

but at a singular null point we have also from above that:

$$2 - 2\cos(\varphi) - \varphi\sin(\varphi) = 0 \qquad (37)$$

and therefore

$$K_{ret}(t,t') = 1 - \beta \sqrt{\sin(\varphi)} / \sqrt{\varphi} \qquad (38)$$

but also at a singular null solution we have from above

$$\beta = \sqrt{\varphi / \sin(\varphi)} \qquad (39)$$

and therefore $K_{ret} = 0$ at a singular null solution.

Therefore, at the singular values of $\beta$, two roots to the null condition merge into one as the singularity is approached from above. Below the singularity, the roots disappear altogether. As the singularity is approached, the two $K_{ret}$ values both approach zero but with opposite sign, and the forces due to each of these sourcepoints on the testpoint approaches infinity, but again with opposite sign. The net force on the testpoint is thus the difference between two large numbers, and numerical results suggest that the behavior of the net force in a neighborhood of a singularity is a fractal type of function of the velocity $\beta$ showing extremely complex behavior and changing sign many times in a very small interval. Three different computer programs were used to analyze this behavior. The first two used 64 bit double precision variables, but it was found that this was inadequate precision for this problem, although they did show fractal behavior. The third program used the extended precision capabilities of Mathematica, and the results obtained were stable to further increases in the working precision of the program. The results presented in the singular neighborhood were done with a working precision of 300 decimal places.

## 6. Summary of Numerical Results

The numerical calculations take a tachyon which is constrained to move on an exact circle with constant speed, and they calculate the radiative self force on the particle. Both Feynman-Wheeler electrodynamics and causal electrodynamics give the same radial force, but for Feynman-Wheeler electrodynamics the azimuthal force is zero.

The first general feature that one can see from Figure 3 is that the radial force is repulsive for most velocities, and that it's magnitude is becoming stronger as the velocity tends to the speed of light or increases towards inifnity. There are discontinuities in the repulsive force, and these occur at the discrete velocities of the previous section (see Table 1).

In the neighborhood of the discontinuities, the repulsive force is a very complex fractal-like function of velocity as is illustrated in Figure 4 for the first singular velocity in Table 1. The sign of the force in this neighborhood can be either repulsive or attractive, and the magnitude can be very large. Finer and finer sampling of the force shows even more structure, like a fractal curve.

The calculations were done with a numerical precision of 300 decimal places using Mathematica, and the results were tested for stability as the working precision was increased and they were stable. The calculation was originally coded using standard 64 bit floating point variables, but this level of precision was insufficient to do the calculation since there is a subtraction of two large numbers resulting in the net force. This necessitated a switch to Mathematica. Thus it can be concluded that for special discrete velocities the tachyon experiences an attractive force (positive values of Z), but this happens only in a neighborhood of the singular velocity values. In this neighborhood the force oscillates rapidly between attractive and repulsive as illustrated in Figure 4.

For causal electrodynamics the azimuthal force is numerically found to be nonzero. The ratio of the azimuthal to the radial force is positive for all the points calculated as illustrated in Figures 5 and 6. When the radial force is attractive, the azimuthal force is opposite in direction to the particle's velocity, and so it is tending to reduce the energy of the particle and can be interpreted as energy lost to radiation. But when the radial force is repulsive, the azimuthal force is in the same direction as the velocity and so it is tending to increase the kinetic energy of the particle. If we imagine a situation where only the radius of the orbit is constrained but the velocity is free to vary, we might get some insight into this situation. Due to the azimuthal force the tachyon in this case would gain energy, and it would consequently slow down until it reached one of the singular velocities, and at that point the force would fluctuate between positive and negative values possibly trapping the particle's velocity at the singular value.

Obviously, helical motions may not be the only confined motions of tachyons in this theory, but they are probably the simplest. It's quite plausible that the tachyons will exhibit chaotic behavior in more general solutions. It's also possible that the number of times that the tachyon's world line crosses the light cone will be proven to be either an invariant of the tachyon's motion or a monotonically increasing function of time. This is because of the singular force that the tachyon will experience whenever two light cone crossing points coalesce into a single point which happens whenever the number of light-cone crossings changes.

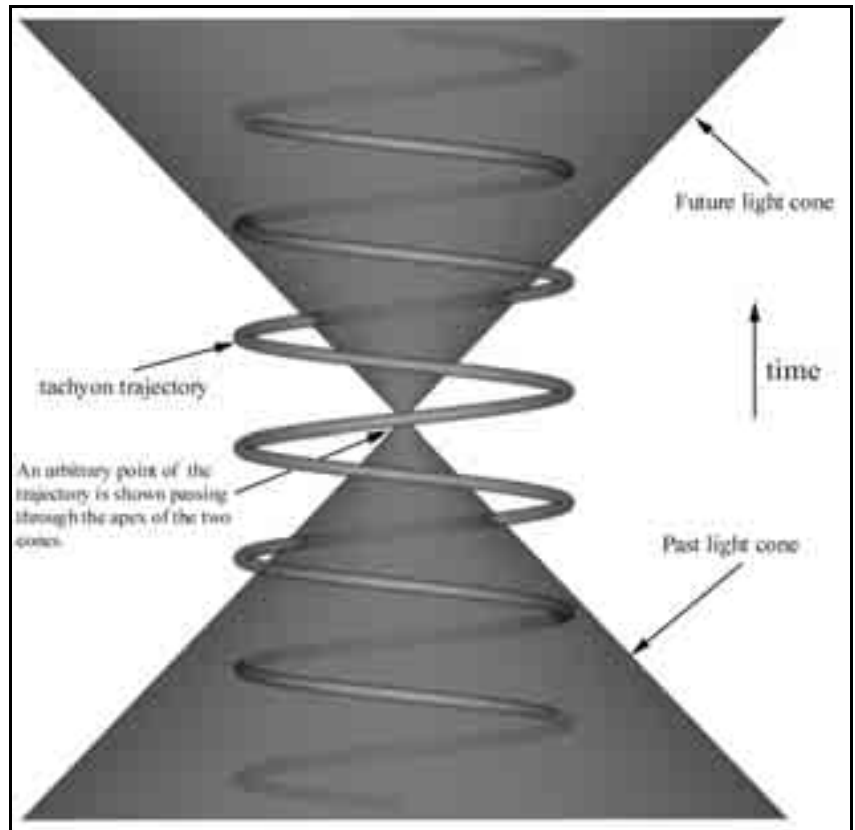

**Figure 1**  A tachyon moving in a helix and intersecting its own light cone

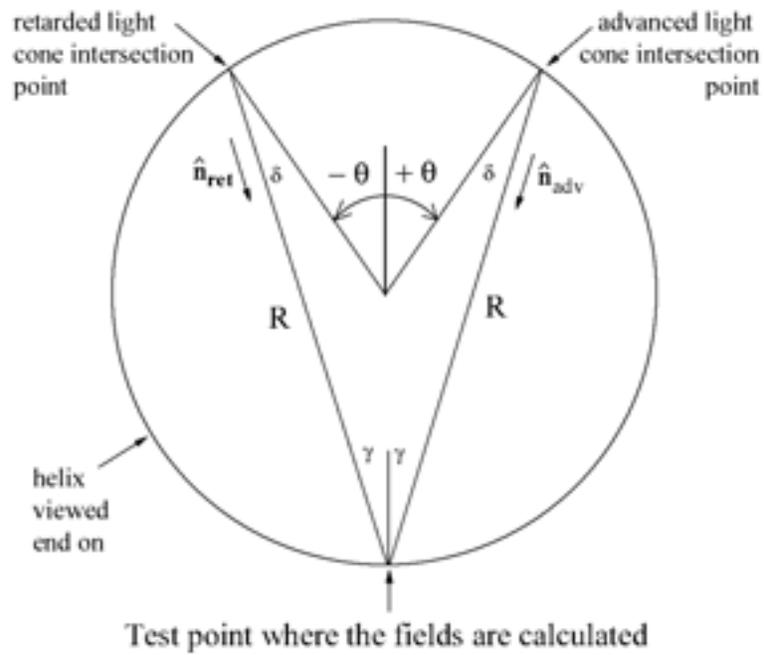

retarded light cone intersection point

advanced light cone intersection point

$\hat{n}_{ret}$ $\delta$ $-\theta$ $+\theta$ $\delta$ $\hat{n}_{adv}$

R R

$\gamma$ $\gamma$

helix viewed end on

Test point where the fields are calculated

**Figure 2** Light cone intersection points come in symmetrical pairs, one point the intersection with the retarded light cone and the other the intersection point of the advanced light cone.

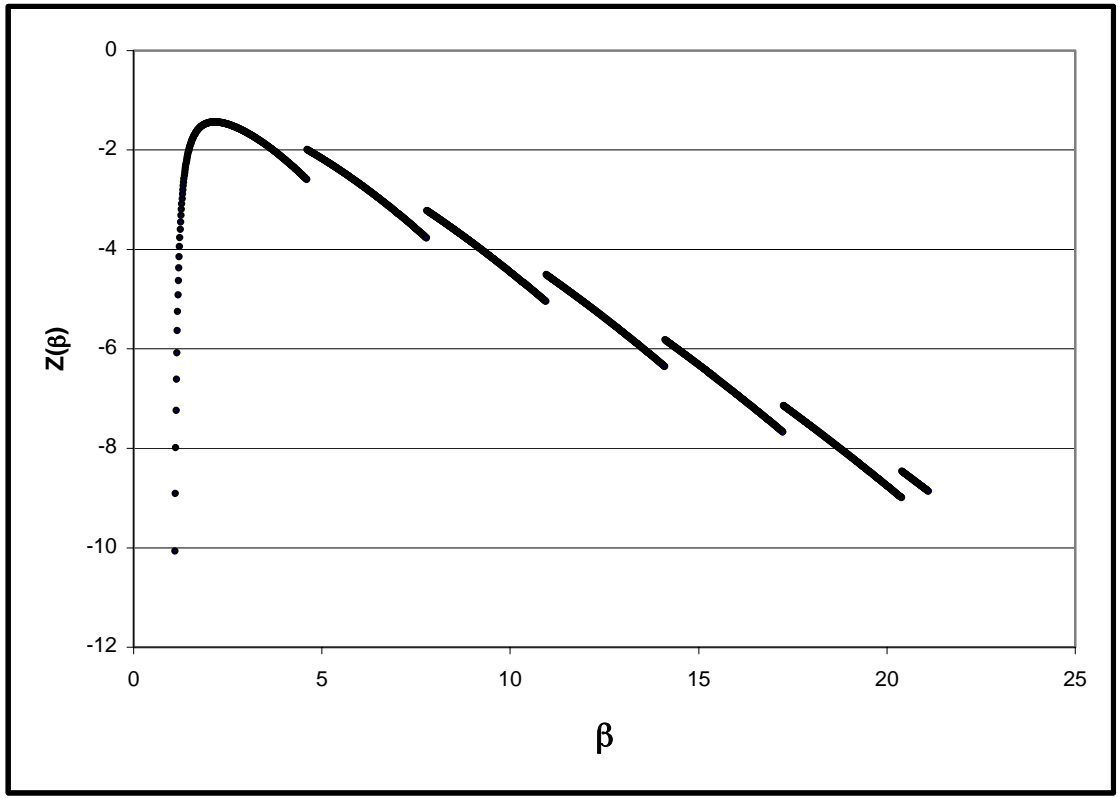

**Figure 3** Coarsely sampled Plot of Z(β) vs. β from 0 to 21. The discontinuities occur at the singularities in Table 1. Finer sampling reveals complex structure in a neighborhood of these singular β values as illustrated in Figure 4.

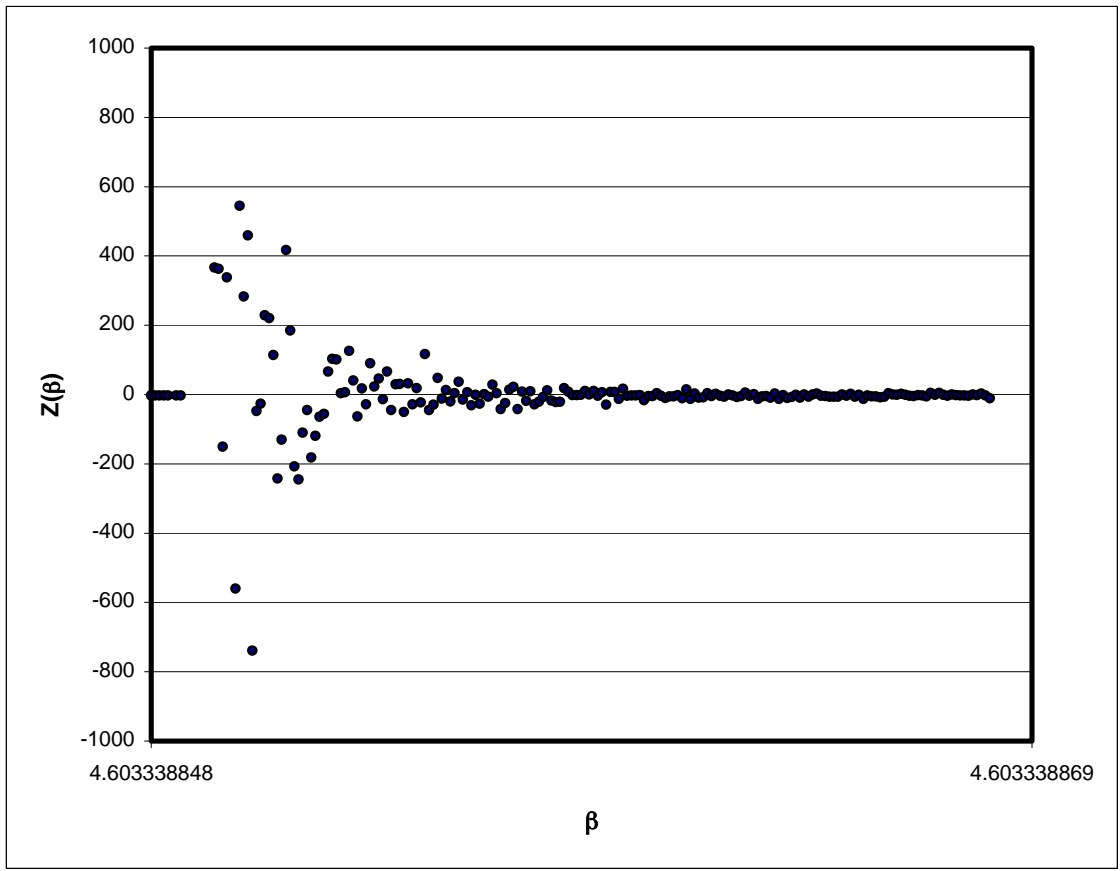

**Figure 4** A finely sampled Plot of Z(β) vs. β in the neighborhood of the first singularity showing positve values of Z corresponding to a net attractive radial force for some values of β.

## 7. The Causal Form for Electrodynamics

When considering the causal form for electrodynamics, the only change from the Feynman-Wheeler case is that the retarded potentials must be used in the field expressions instead of the average between the retarded and advanced potentials. We consider a particle which is constrained to move in a helix, and we calculated the electromagnetic force on it. It can be shown that the radial force is identical to the Feynman-Wheeler case but the azimuthal force is now nonzero. It is convenient to express the azimuthal force in terms of a ratio with the radial force. Figures 5 and 6 show plots of the calculated azimuthal forces for two ranges of β. These plots show that the sign of the ratio is always positive. When the radial force is attractive in sign, the azimuthal force is opposite in direction to the particle's velocity. This apparently represents radiative reactive force due to energy lost to radiation. It should be noted that it is well established [6,13] that as a tachyon loses kinetic energy, its speed increases.

$$F_{azimuthal} = F_r \varepsilon(\beta) \tag{40}$$

We see from Figure 5 that as the velocity increases, the azimuthal force becomes smaller as compared to the radial force.  This means that in the limit of infinite velocity the tachyon will not radiate away energy.

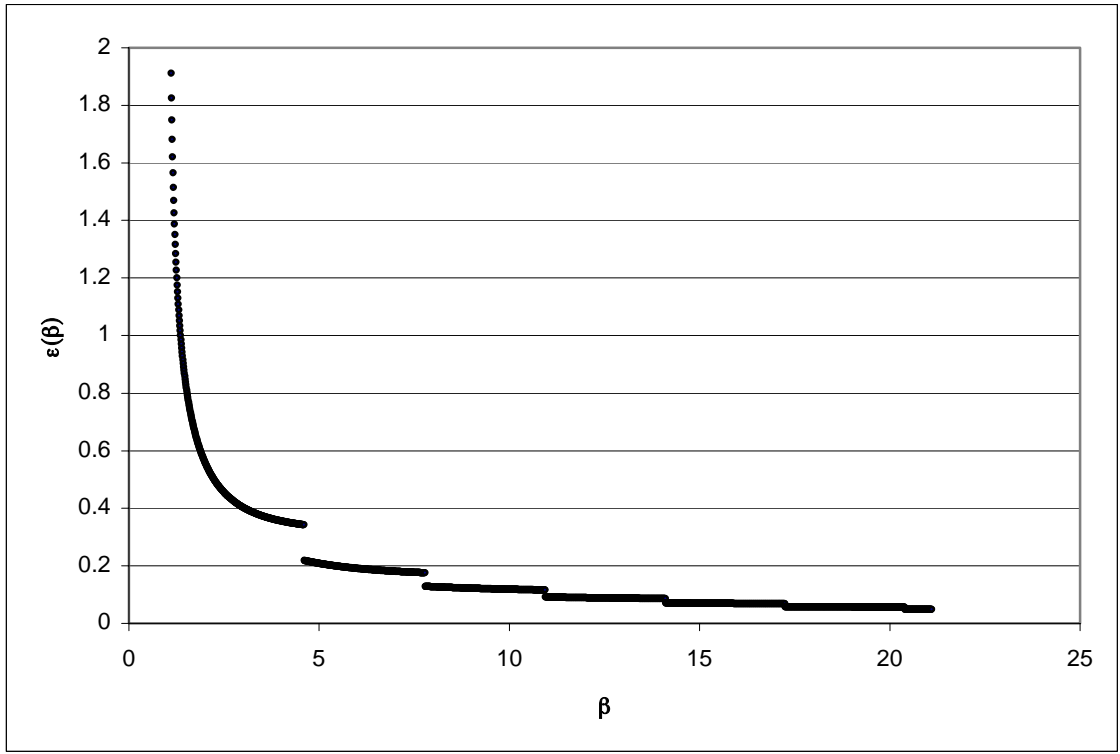

**Figure 5**  The ratio of azimuthal to radial force with causal (retarded) electrodynamics. The radial force is the same as for Feynman-Wheeler electrodynamics.

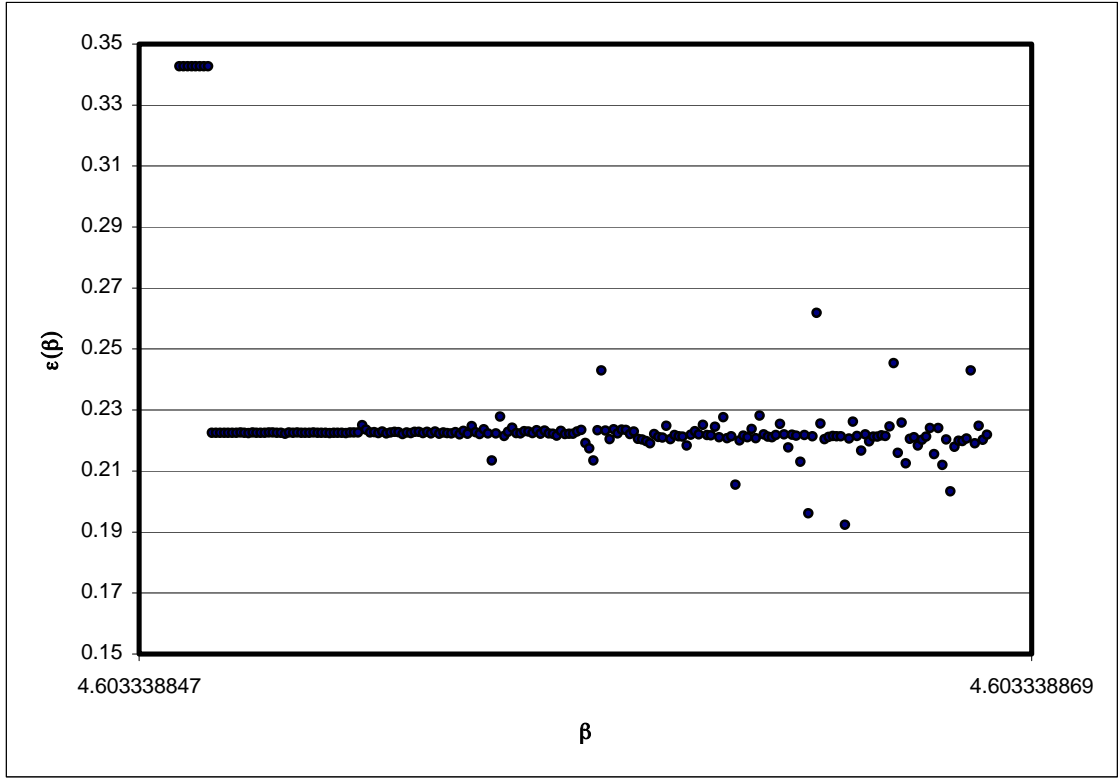

**Figure 6** Finely sampled ratio of Azimuthal to Radial force with Causal (retarded) Electrodynamics

## 8. Tunneling of Classical Tachyons

One can see quite easily that something like tunneling must occur for classical tachyons. Consider such a particle moving in one dimension towards a classical turning point where the kinetic energy would vanish. One can easily derive the kinetic energy of a tachyon from the momentum expression by using the fomula:

$$\mathbf{P} = m_0 \mathbf{v} / \sqrt{\beta^2 - 1} \tag{40}$$

$$E_k = \mathbf{P} \cdot \mathbf{v} - \int \mathbf{P} \cdot \dot{\mathbf{v}} dt = \frac{m_0 c^2}{\sqrt{\beta^2 - 1}} \tag{41}$$

where $E_k$ is the kinetic energy. As the kinetic energy for the particle approaches zero, its velocity must approach infinity in the direction into the forbidden region. So in the next instant the particle will move beyond the classical turning point, and thus tunneling is inevitable. We may illustrate this mathematically as follows. Consider integrating equation 1 with the potential energy and initial conditions shown in figure 7.

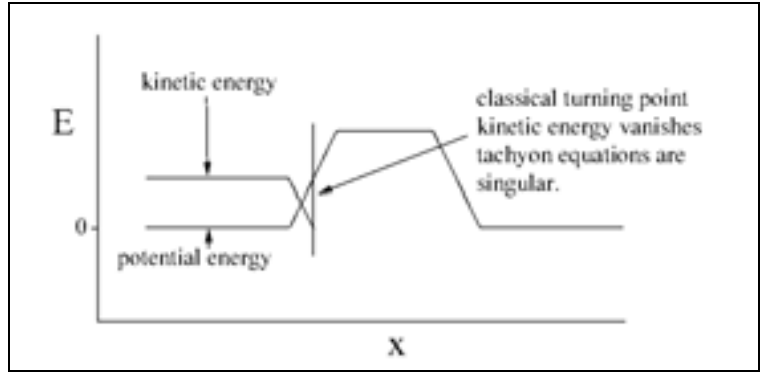

**Figure 7**. Energy diagram for tachyon tunneling

at the classical turning point (CTP) the following limiting results are reached

$$\lim_{x \Rightarrow CTP} E_k = 0; \quad \lim_{x \Rightarrow CTP} \mathbf{P} = m_0 c \frac{\mathbf{v}}{|\mathbf{v}|}; \quad \lim_{x \Rightarrow CTP} \beta = \infty \qquad (42)$$

Attempting to integrate the force equation (1) through the turning point leads to a problem since the momemtum will become smaller in magnitude than $m_0 c$ which is impossible for a tachyon. Therefore if the theory is to be seriously considered, some way of handling classical turning points must be found. There appears to be only one way to include such classical turning points in the tachyon theory. The tachyon must begin to travel backwards in time after passing through a turning point. This is suggested by the fact that the velocity - being infinite at the classical turning point - can turn backwards in time at that point without a discontinuity. Moreover, it is well known that Lorentz transformations can make any point in a tachyon trajectory have infinite velocity, and this will in many cases cause the trajectory to reverse direction in time either before or after the point where the velocity is infinite. Figure 8 illustrates the tunneling behavior in this circumstance. The trajectory is a splice of the incident tachyon's motion forward in time, a splice of backward time motion in the forbidden region, and again forward motion on the other side of the barrier. Obviously, this type of behavior appears to violate causality. If the tachyon is initially moving very fast, then the traversal through the barrier will occur essentially instantaneously, and it will tunnel out the other side essentially at the same time it entered.

When the tachyon is incident on a repulsive potential at an angle, tunneling may or may not occur depending on this angle and the other dynamical parameters of the motion. Figure 8 illustrates this feature. If a tachyon is moving in a helix and especially if there is some chaotic motion superimposed on the tachyon's helical motion, then the angle of attack that the tachyon hits a barrier will be quite random and consequently the likelihood of transmission or reflection will be governed by probabilistic laws, similar to quantum mechanics.

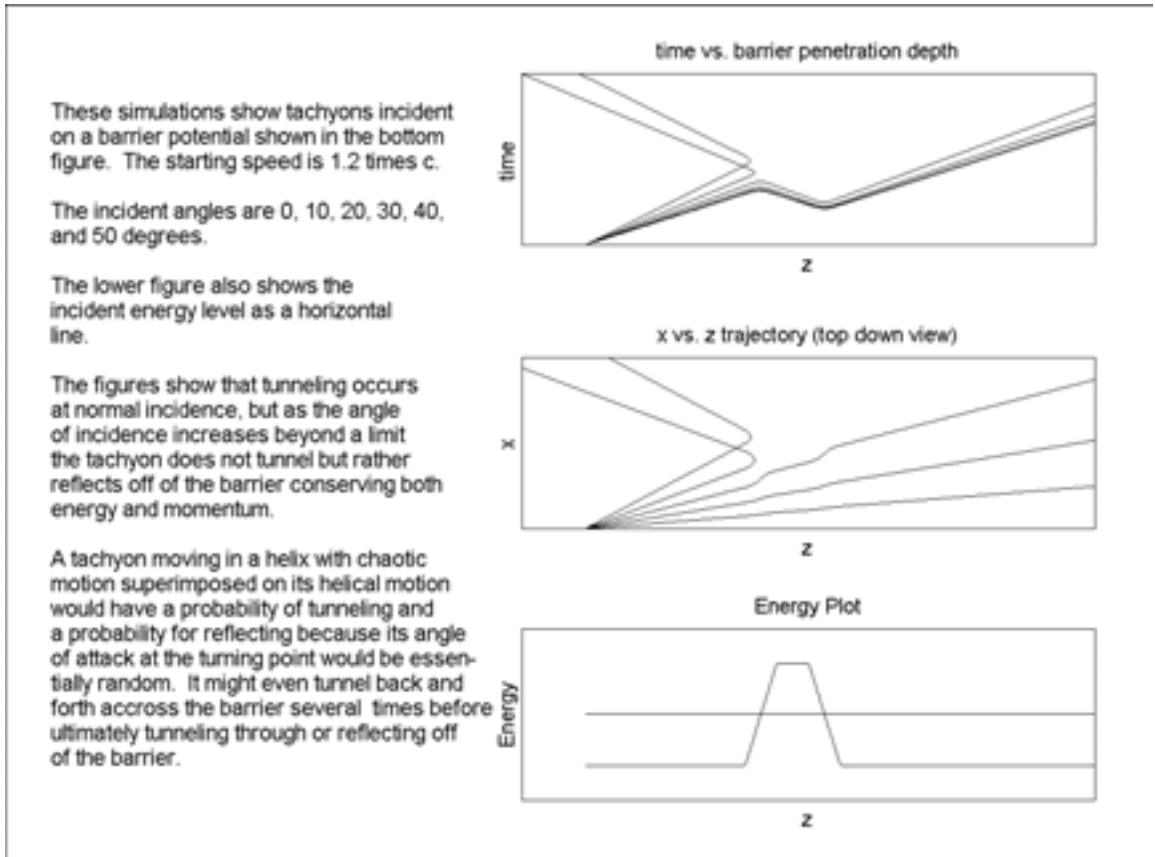

**Figure 8.** Simulations of a tachyon tunneling through a barrier

## 9. Ramifications

Theoretical conjectures are presented in the next few subsections. They present ideas which are of a hypothetical nature that serve to highlight the potential of this subject.

### 9.1 Radiation and Radiative Reactive Forces

The results of the previous section show that when a tachyon crosses its own ligthcone it can experience a radiative force which causes its energy to change. This case was not considered by Ey and Hurst [5] who showed that tachyon's do not experience the same local radiative self-force as do bradyons (slower than light particles) in the Lorentz-Dirac equation. Ey and Hurst concluded from this that tachyons can never radiate. However, the above result shows that this conclusion must be amended. Tachyons can still radiate when they cross their own light cone. The Ey and Hurst result is still extremely important because it shows that the classical equations of motion for tachyons do not have the radiative reaction term of the Lorentz-Dirac equation. As a consequence, the Tachyon equation of motion does not appear to have the same problems with runaway solutions as the Lorentz-Dirac equation for bradyons. It's interesting to note that the runaway solutions of the Lorentz-Dirac equation involve cosh and sinh functions of the proper time for the time and one spatial coordinate [17]. This is mathematically equivalent to a circular motion which has been analytically continued to imaginary values of proper time and where one space coordinate has become timelike as a result of the superluminal transformation. Tachyons may be considered to have imaginary values of proper time, and so perhaps there is a geometrical relationship between the runaway solutions of the Lorentz-Dirac equation and the helical motions that we have found here. Perhaps the runaway solutions are actually required for extended lorentz invariance where superluminal boosts are allowed [6,7]. Perhaps they are just the helical motions that we are describing here, but viewed from a superluminal frame in which the tachyon is moving slower than light.

## 9.2 Causality

Tachyons, if they exist, may always or almost always be bound up in helical motions similar to the solutions exhibited here or in other confined orbiting motions. Then they would appear to be moving slower than light and to have intrinsic angular momentum. If this were the case, then the tachyons couldn't be used to send faster-than-light signals, and so the causality paradox would be simply resolved.

## 9.3 Relevance to hidden variable theories and stochastic models of quantum mechanics

The results here are applicable in the hidden variable effort of quantum mechanics. The hope that hidden variables may be the way of nature continues to attract prestigious adherents such as Gerard t'Hooft [18]. When one surveys the field though, one finds that despite great effort, a satisfactory derivation of quantum mechanics from a hidden variable perspective still eludes researchers.

First in the list of problems which must be overcome is Bell's theorem [19-21] which states that a hidden variable model must have nonlocality or superluminal connections built into it. Bell's theorem is usually thought to be the most decisive objection to hidden variable models. Tachyons certainly provide for the possibility of nonlocality. But then the tachyon theory must somehow end up being causal. Tachyons bound up in helical motions (or other more complex confined orbital motions) will be effectively causal.

It is a long-standing problem to include spin into hidden variable models, particularly in a relativistically covariant way. Again, the helical motion of tachyons is a natural candidate for spin. In fact, Recami and Salesi [22] have recently suggested that classical particles with spin may be the key ingredient in finding an explanation of the quantum mechanical potential as it appears in Bohm's hidden variable theory [23 and references therein]. The current theory is fully relativistically covariant, and an additional benefit is that there are apparently no runaway solutions plaguing this theory as in the Lorentz-Dirac equation for charged bradyons.

With the Feynman-Wheeler interaction form the tachyon doesn't radiate when it moves in a helix, but with the causal interaction form energy is lost by the tachyon to radiation if the force is attractive. If the tachyon has more complicated confined solutions, it is possible that radiationless states may someday be found. Radiation free states may have very high velocities because as the tachyon radiates energy away its speed increases and the energy loss for an orbiting tachyon decreases with speed (as can be surmised from Figure 5). It's very likely, owing to the complex nonlinear equation of motion of the self-interacting tachyon, that if the simple helical motions studied here are perturbed slightly then the tachyons will exhibit chaos and this could perhaps be a model of quantum indeterminism. Some numerical investigations of this possibility have been carried out but results thus far are inconclusive. The singular force requires a very fine time sampling and makes the calculations difficult.

A quantum mechanical wave might naturally be visualized as the confined motion of a tachyon which moves ergodically in a short time to fill out the wave. This confined motion would be a superposition of helical motion with chaotic motion of the center of the helix moving rapidly around the wave's region of support. The wave function could collapse almost instantaneously in this picture because of the tachyon's great speed. Such chaotic motion might provide the stochastic behavior in hidden variable models like stochastic quantum mechanics (Nelson[26], Davidson[ 27-28], and references therein) . Indeed, tachyons are much better suited for a description of stochastic models of this type than are bradyons because of the singular nature of the Markov process used in these models and their singular velocities. A single tachyon acts in some ways like a many body problem because of its ability to interact with its own past positions at possibly a large number of points. This is a natural explanation of wave-particle duality. In a two slit diffraction experiment for example, the tachyon would have the opportunity of going through both slits many times. This avoids the usual arguments against the particle's ability to "know whether or not the other slit was open" when it passes through one of the slits.

Another approach to explore in looking for a hidden variable model would be to postulate the classical zero-point vacuum radiation model of stochastic electrodynamics [24,25] and then analyze how the tachyon would diffuse in this background. The motion would be random due to the interaction with the background radiation, and also could be chaotic from its own self-interaction.

The Aharanov-Bohm effect [29] is another perplexing phenomenon that needs to be explained by hidden variable models. In the analysis above it was noted that singularities occur in the field expressions when the Cerenkov cone from a retarded source charge points to the test point. These singularities have no analog for slower than light particles. If the electrons in a solenoid were actually tachyons moving in helices, then they would produce a great multitude of singular field points even outside the solenoid. These might be hard to detect in most situations, as they might be randomly oriented and quickly changing with time and in most cases they might add up to zero net force on moving particles outside the solenoid. But maybe the Aharanov-Bohm experiment is a case where the effect can be measured. The point is that the Cerenkov singularity is a mechanism for something real to penetrate the region of the solenoid and affect a particle outside of it. Perhaps it can explain the Aharanov-Bohm effect.

The helical solutions that we have found may behave when perturbed in some ways like relativistic strings when their velocity tends to infinity. A circular motion will map out a helix in space time and if the velocity of the tachyon is very high, then the pitch of the helix will be small, and the tachyon's worldline will look like a cylindrical surface. Its mechanical behavior may begin to resemble that of a relativistic string in this limit.

Another problem for a hidden variable theory is how one is to interpret the double-valued nature of the half integral spin variable. Perhaps this too has an interpretation in the present theory. If one considers superluminal boosts added to the Lorentz group as is done in the extended relativity [6,7], then maybe it is possible to rotate through 360 degrees by doing it in the superluminally boosted frame, and then come back to the original unboosted frame. In some circumstances the classical coordinates may not return to their original values in this case. The reason is that the expression for proper time has a square root singularity in the velocity variable. Superluminal boosts may involve an analytic continuation in the proper time variable around a square root singularity which could yield the double-valued nature of spin in a classical albeit tachyonic framework. The double-valued nature of half-integral spin might be a result of the double-valued nature of the square root singularity for proper time when analytically continuing it to superluminal frames. It's plausible that when the tachyon circles around the axis of spin, then the spin is double-valued but when the helix itself circles slowly around an axis, it generates a single-valued angular momentum analogous to orbital angular momentum in quantum mechanics. We defer consideration of this idea to a future publication.

So we have a possible way around several of the most disconcerting objections that have been raised against hidden variable theories. Even it some of these conjectures prove incorrect, they can serve as a guide for those who are seeking a hidden variable description of quantum mechanics. If one is so inclined, it's extremely difficult to resolve even one of the above paradoxes in a classical theory. The present theory potentially has a resolution for all of them. This is truly remarkable and unique. Moreover, from the point of view of Occam's razor, the present tachyon theory is very economical. It posits only pointlike charged particles. What other explanation for the existence of spin could be this simple in a hidden variable theory based on classical physics?

It is plausible that chaos will be observed in a more general solution to the tachyon equations of motion. It's also plausible that the number of times in the past that a tachyon intersects it's own light cone will turn out to be a constant of the motion or a monatonically increasing function of time, because in order for this number to change, the tachyon must experience a singular force corresponding to $K_{ret}$ vanishing as the tachyon's trajectory becomes tangent to its own light cone. If this were true the tachyon could never straighten out and move in a straight line.

It is very likely that when one analyzes two or more tachyons, that circular solutions to their motions will be found in which the group of tachyons move together in a circle.

**9.4 Can the Fine Structure Constant be Derived?**

Edward Nelson postulated that the interaction of the classical electromagnetic field with point charged particles may lead to chaotic behavior when treated correctly from a mathematical point of view ([26] page 65). If the current theory exhibits chaos when the helical motions of tachyons is perturbed, then it will be a confirmation of Nelson's conjecture. As Nelson points out, electromagnetism is the natural theory to look for a chaos-based model of quantum mechanics because the fine structure constant $\alpha$ is dimensionless.

$$\alpha = \frac{e^2}{\hbar c} = 7.29720 \times 10^{-3} = 1/(137.0388) \tag{43}$$

If the mean radius of a perturbed helical motion could be calculated, then the angular momentum of the orbiting tachyon could be equated to either $\hbar$ or to $\hbar/2$ (it's not clear which of these choices is correct) and this would yield a number for $\alpha$ which could be compared with the experimental values in (43). It is not a trivial matter to find the mean radius though as can be seen from Figure 4 which shows how sensitive the attractive force is to the velocity of the motion. Table 1 shows that the current theory has already yielded a number of dimensionless constants and that it is calculable. Nevertheless the fine structure constant has thus far proved elusive.

Another attempt to derive the fine structure constant entirely within the context of electromagnetism is the postulate that it is a value which renders quantum-electrodynamics finite [30,31,32]. This postulate has so far not been confirmed.

**9.5 Multiple Tachyon Orbits**

It seems certain that there will be helical orbiting solutions involving more than one tachyon with all the tachyons moving in the same circle with the same speed. The same types of singularities will occur when the testpoint is on the Cerenkov cone of the sourcepoint as occur for a single tachyon. This will happen at discrete eigenvalues, but the values will be different than the ones found here for a single tachyon. Qualitatively the same singular fractal-like force (Figure 4) will clearly be found for multiple tachyon orbits, and this force could in some cases prevent the particles from escaping their orbits just as it may prevent the single tachyon from ever straightening out its orbit and flying in a straight line. It's natural then to try and see if some variant of the quark model could be understood in this framework. The singular Cerenkov force is a natural mechanism to explain quark confinement.

# 10. Conclusion

The mathematical theory of self-interacting charged tachyons is complex and intriguing. Spin and tunneling emerges from the theory, and unusual singular forces act on tachyons that have no analog in the rest of classical physics. The singular forces occur when the testpoint lies on the Cerenkov cone of the source point. Perhaps a useful hidden variable theory of quantum mechanics can be developed with charged classical tachyons as the central agents. Many paradoxes or difficulties are conceptually overcome if tachyons in helical orbits could be the hidden variable models for elementary particles. Charged tachyons may be masquerading as slower than light particles because they are moving in tight helices that disguise their faster-than-light motion. These particles would appear to have intrinsic angular momentum (spin) and magnetic moment, even though the actual charged tachyon had no spin.

If a tachyon moves in a helix, then the center of the helix will transform as a slower than light particle under Lorentz transformations, and therefore it will always move forward in time on the average no matter what frame it is viewed in. This could largely mitigate the causality problem for tachyons whereby they could be used to send signals backward in time as in the Tolman Paradox [8]. If they move in tight helical motion, they can't easily be used to signal backwards in time.

Perturbing the tachyon motion from a perfect helix will very likely lead to chaos, and a careful analysis of this chaotic behavior in simulation may someday yield the fine structure constant.

## Acknowledgements

The author acknowledges extensive discussions with Mario Rabinowitz, useful conversations with Vladimir Kresin, and helpful correspondence with Erasmo Recami, Angas Hurst, and Edward Nelson.